\documentclass[authoryear]{article}

\usepackage{cite}
\usepackage{amsmath,amssymb,amsfonts}
\usepackage{algorithm}
\usepackage{textcomp}
\usepackage{setspace}
\usepackage{tikz}
\usepackage{a4wide}

\usepackage{url}
\usepackage{bm}

\usepackage{graphicx}
\usepackage{color} 
\usepackage{todonotes}

\newcommand{\change}[1]{\textcolor{black}{#1}}
\newcommand{\DP}[1]{}
\newcommand{\MG}[1]{}
\newcommand{\SB}[1]{}

\title{Time Aggregation Techniques Applied to a Capacity Expansion Model for Real-Life Sector Coupled Energy Systems}
\author{Mette Gamst$^1$, Stefanie Buchholz$^2$, David Pisinger$^2$ \footnote{Corresponding author - email: dapi@dtu.dk}\\
$^1$ Energinet, Tonne Kj\ae{}rsvej 65, 7000 Fredericia, Denmark\\
$^2$ DTU Management, Akademivej 358, 2800 Kgs.Lyngby, Denmark
}
\date{\today}

\begin{document}

\maketitle

\section*{Abstract}
Simulating energy systems is vital for energy planning to understand
the effects of fluctuating renewable energy sources and integration of
multiple energy sectors. Capacity expansion is a powerful tool for
energy analysts and consists of simulating energy systems with the
option of investing in new energy sources. In this paper, we apply
\change{clustering based} aggregation techniques from the literature to very
different real-life sector coupled energy systems. 
We systematically compare the aggregation
techniques with respect to solution quality and simulation
time. Furthermore, we propose two new \change{clustering} approaches with
promising results.  We show that the aggregation techniques result in
consistent solution time savings between 75\% and 90\%. Also, the
quality of the aggregated solutions is generally very good. To the
best of our knowledge, we are the first to analyze and conclude that a
weighted representation of clusters is beneficial. \change{Furthermore, to the best of our knowledge, we are the first to recommend a clustering technique with good performance across very different energy systems: the k-means with Euclidean distance measure, clustering days and with weighted selection, where the median, maximum and minimum elements from clusters are selected.} A deeper analysis
of the results reveal that the aggregation techniques excel when the
investment decisions correlate well with the overall behavior of the
energy system. We propose future research directions to remedy when
this is not the case.

\noindent
\textbf{Keywords:} Capacity expansion; energy system models; time aggregation; clustering; solution time reduction



\section{Introduction}
\label{intro}
Simulating energy systems is vital for energy planning. The green
transition demands increasing introduction of fluctuating renewable
energy sources and integration of multiple energy sectors. Simulations
are necessary to understand the behavior in such sector coupled energy
systems. Capacity expansion consists of simulating energy systems with
the option of investing in energy sources. This is a very powerful
tool to understand the effects of new technologies.

\change{A tremendous amount of work in the literature considers solution methods for the capacity expansion problems but focuses on single methods or specific energy systems, see e.g. \cite{brown2018,2,Babatunde2019,sadeghi2017}. The contribution of this paper is to recommend a solution approach with overall good performance. It does so by comparing existing and new solution methods on very different energy systems. This differs from the more common approach from the literature of solving the capacity expansion problem for specific energy systems to perfection. This is further elaborated in the literature review.}

In this paper, a capacity extension model consists of a year in
one-hour resolution, i.e. of 8760 hours. The underlying energy system
may consist of many areas (e.g. geographical areas or spot
market bidding zones) and energy types (e.g. power, district heating,
gas), which results in large capacity extension models. Solving the
NP-hard capacity expansion problem is thus time consuming and at times
even intractable. Aggregating time steps is a common method to
simplify simulations to reach tractability. The literature suggests a
wide variety of aggregation techniques. Many studies,
however, consider \change{specific} systems \cite{Babatunde2019} and only few contributions compare their results with the literature \cite{2}.

The novelty of our work lies in analyzing the effect of time
aggregation methods on the real-life sector coupled energy systems. To
the best of our knowledge, we are the first to apply and compare
multiple time aggregation methods on significantly different energy
systems. This provides much insight in the potential of time
aggregation techniques without the risk of overfitting the methods to
specific energy systems.  
Furthermore, we propose two new aggregation methods with
promising results. All time aggregation techniques are based on
clustering. We analyze methods for selecting cluster representatives
and
conclude that weighted selection has superior performance. Finally, we
provide a deeper analysis of the achieved results to highlight
interesting future research areas in time aggregation techniques.  The
paper addresses gaps in the current literature of time aggregation
techniques applied to capacity expansion models: comparison of methods
on very different energy systems, and analysis of selection strategies
in the clustering methods.

The paper is structured as follows. The literature review in Section \ref{lit} introduces the capacity expansion problem, considers aggregation techniques from the literature and concludes with a motivation for the contributions of this paper. The considered aggregation techniques are presented in Section \ref{sol} and the real-life energy systems in Section \ref{test}. The clustering methods are evaluated on the energy systems in Section \ref{res}. The evaluation leads to the proposal of two new aggregation techniques in Section \ref{new}. Section \ref{fur} contains a discussion of lessons learned and of possible future work. Finally, conclusions are drawn in Section \ref{conc}.

\change{
\section{Literature review}
\label{lit}
The literature review first focuses on the capacity expansion problem and afterwards on solution methods for the capacity expansion problem.
}

\change{\subsection{The Capacity Expansion Problem}}
\label{cep}
\change{
We first consider the Unit Commitment Problem (UC) and then the Capacity Expansion Problem (CEP).
The importance of considering the UC together with the CEP has been quantified in several studies, which show that ignoring the operation details from the UC leads to underestimations in flexibility measures and overestimation of base-load and 
renewable energy sources (RES) \cite{nogales2016,hua2017,deane2012}. 
}

\change{
The Unit Commitment Problem simulates an energy system, where demand
must be met every hour \cite{baldwin82}. It determines the operation schedule of
generating units, subject to demand, technical constraints, available
RES, etc \cite{Mallipeddi27}. It is NP-hard and includes 
binary decision variables for turning production units on and off \cite{3}.  A general mathematical model for the unit commitment problem can be summarized as:}
\change{
\begin{align}
\text{min}\quad	& \text{total system costs}  \label{genobj}\\
\text{s.t.}\quad&	\text{production + import + storage discharge = }  \text{demand + export + storage charge}\label{genbal}\\
&\text{physical constraints on production units}\label{genpu}\\
&\text{available RES}\label{genres}\\
&\text{storage and electric vehicle constraints}\label{gensto}\\
&\text{capacity on interconnection lines}\label{genicl}
\end{align}}
\change{
The total system costs \eqref{genobj} include fuel prices, costs on production and
fuel consumption (e.g. taxes, subsidies), emission
costs, import and export costs, operation and management costs,
startup costs and costs associated with flexible demands. Balance
constraints \eqref{genbal} ensure that supply and demand meet in every area and in every hour. Supply consists of
production, import and storage discharge. Demand consist of demand, export and storage
discharge. Balance constraints include slack variables, which are
activated when demand and supply cannot be matched.  Physical
constraints on production units \eqref{genpu} include unit commitment, fuel
consumption, efficiency, startup costs, technical production limits,
production technology (condensation, back pressure, extraction,
turbine bypass, heat boiler, etc.) and ramping.  Constraints \eqref{genres} consider available RES subject to curtailment options. Storage and electric
vehicle constraints \eqref{gensto} include capacities, losses, charge and discharge
rates, and, for electric vehicles, charging before driving. Finally, \eqref{genicl} ensure that capacities on interconnection lines are satisfied.}

\change{
The UC is subject to a considerable amount of research in order to increase the tractability of the problem by e.g. linearizing non-linear constraints \cite{ana2013}, reducing the number of binary variables \cite{carrion84}, tightening the formulation \cite{ostrowski85}, applying dynamic programming solution methods, heuristics and math-heuristics, see  \cite{Mallipeddi27} for a survey and comparison of methods.
}
~\\

The Capacity Expansion Problem (CEP) extends the Unit Commitment Problem
with investment decisions. This enables analyses of introduction of
new technologies, energy mix in case of rapid technology development,
flexibility needs, etc. \cite{2}.
\change{
A general mathematical model for the CEP is summarized as: }
\change{
\begin{align}
\text{min} \quad	& \text{total system costs \textbf{+ investment costs}}	\label{geninvbalobj}\\
\text{s.t.} \quad	&\text{production + import + storage discharge = } \text{demand + export + storage charge} \label{geninvbalbal}\\
&	\text{physical constraints on production units}\label{geninvbalpu}\\
&	\text{available RES}\label{geninvbalres}\\
&	\text{storage and electric vehicle constraints}\label{geninvbalsto}\\
&	\text{capacity on interconnection lines}\label{geninvbalicl}\\
&\textbf{utilization $\leq$ investment}\label{geninvbalinv}\\
&	\textbf{min investment $\leq$ investment $\leq$ max investment}\label{geninvbalinvbounds}
\end{align}
The objective function \eqref{geninvbalobj} ensures that an investment takes place when
the savings of utilizing the investment exceed the investment cost. “Utilization” in 
constraints \eqref{geninvbalinv} represents how the investment is utilized in
terms of production (for production units and RES), inventory
level (for storage units) or import and export (for interconnection lines). The constraints say that the investment must be large enough to facilitate the desired utilization. Finally, bounds \eqref{geninvbalinvbounds} ensure that investments are within the user defined bounds.
}

\change{The CEP is widely used for optimizing the configuration of future energy systems. Examples of applications are non-trivial power systems \cite{michaet2015}, integration of renewable energy \cite{oree2017}, large energy systems with sector coupling such as power, gas, transport and heating \cite{brown2018}, and the impact of scenarios \cite{xie2019}. An important type of CEP is  the \textit{Generation Expansion Planning} (GEP) which considers electricity systems \cite{Babatunde2019}. The GEP optimizes the capacities and can give input to locations and commissioning times, see \cite{oree2017,nikolaos2018,sadeghi2017,dagoumas2019} for reviews and for applications on different energy systems.}

\change{
As mentioned, incorporating the UC in the CEP is essential to analyse the need for flexibility capacities and the integration of RES \cite{nogales2016,hua2017,deane2012}. 
This is particularly the case in systems with much RES \cite{shortt2013,palmintier2015} or when alternative flexibility sources are analysed \cite{poncelet2019}. Recall that the UC is NP-hard. Several studies have thus considered simplified approaches to include the UC in the CEP
 \cite{hua2017,carrion84,ostrowski85,ana2013}, e.g., by only considering a subset of the UC constraints  \cite{jonghe2011,batlle2013,manuel2014,hans2018}, however,
 the resulting problem still remains very difficult to solve. 
 Further research has looked into sophisticated solution approaches such as Benders decomposition \cite{anna2018} and Dantzig-Wolfe decomposition \cite{Quiroz2016}. A very popular approach is time aggregation, where only a subset of the 8760 hours of the year is solved. This reduces the size of the problem to make it more tractable, but at the cost of losing precision.
Several literature studies apply different sophisticated strategies to select a subset of the 8760 hours and conclude that the quality is satisfying, 
  see e.g. \cite{michaet2015,Sisternes2015,Poncelet2015,Poncelet2016,scott2019}. The next section further elaborates time selection methods.}

\change{
\subsection{Time aggregation techniques for the Capacity Expansion Problem}
}

\change{
Many different time aggregation techniques exist, spanning from simple heuristic selections \cite{Fripp2012} to optimization methods \cite{Poncelet2016}. Heuristic approaches may be too simple and are at times associated with insufficient capture of variability \cite{Merrick2016} while the optimization approaches suffer from high computational efforts \cite{Poncelet2016}. It seems that a good compromise between quality and computational tractability is achieved by using clustering procedures \cite{kotzur2017,heuberger2017,kotzur2018,abeer2018}. 
A survey along with a proposed classification of time aggregation methods can be found in 
\cite{2}.}

In this article, we focus on clustering methods. This literature study first reviews literature on
clustering methods, then on how to select elements from clusters, then
on comparing methods and finally we discuss how this paper contributes
to closing gaps in the literature.
\bigskip

\subsubsection{Clustering techniques}
The aim of all clustering approaches is to minimize the similarity between clusters while maximizing the
similarity within each cluster \change{\cite{fisher1958}. Clustering approaches differ in how
they group elements into clusters and how they select elements from each cluster.
An example of a clustering technique can be found in \cite{Nahmmacher2016}, which clusters days according to the hierarchical clustering procedure and where the day closest to each cluster centroid is chosen. } Other popular approaches are k-means \cite{kmeansoriginal} and fuzzy clustering \cite{fuzzyclusteringoriginal}.

Clustering techniques can be
categorized into either Exclusive (each element is assigned to only
one cluster) or overlapping cluster techniques (each element is
assigned to all clusters with a degree of membership) \cite{2}. In
relation to time aggregation, most clustering approaches belong to the
Exclusive category, although the use of some overlapping clustering
techniques, such as a fuzzy clustering, also exists \cite{2,14,15}. The most common Exclusive techniques are the Hierarchical clustering and the k-means clustering; the former builds a hierarchy of clusters through a sequence of nested partitions, while the latter initializes a grouping which is then iteratively improved \cite{13}.

\subsubsection{Selecting elements from clusters}
The hypothesis of using clustering techniques for time aggregation is
that the resulting selected elements include all information needed to
make optimal investments. However, perfect similarity within the
clusters is usually not achieved, wherefore the strategy of selecting
elements from each cluster influences the aggregation performance. The
authors of \cite{12} provide an overview of selection strategies including
cluster average, element closest to the cluster average and random
element selection. Average representatives are frequently criticized
for smoothing the profiles \cite{16,Poncelet2015} with a consequence of
underestimating the need for storage capacity and storage technologies
\cite{18}.  On the other hand, a random selection shows good results in \cite{2} where also both minimum and maximum element selections are included in the comparison. Apart from \cite{2}, comparisons of element selections are
seen in \cite{Nahmmacher2016} and \cite{14}. After the selection, the elements are weighted such that the aggregation reflects the relative importance of the
elements in the original problem. Typically, fixed weighting is
applied, assuming each cluster element to be equally important
\cite{15}. The weighting could also choose only to represent a partition of
the clusters \cite{19}. To our knowledge, there is no clear conclusion
regarding the existence of a single best selection criteria nor a
single best weighting strategy.

\subsubsection{Comparison of aggregation methods}
\change{Relatively few articles in the literature compare clustering methods systematically}. In the
following, we bring forward some of the more interesting
contributions. \cite{4} compare clustering procedures, which group days
into clusters. Only single clustering techniques are considered and
each clustering uses a similarity metric based on the principle
components of the day elements. The clustering procedures are the k-means, the fuzzy cmean clustering and hierarchical clustering with
varying linkage criteria. Two different element selections strategies
are considered; the mean and the median element selection. In both
cases, the selected element is weighted (repeated) according to the
number of elements in its cluster. The paper analyzes electricity
demand only, and furthermore it only considers how well the original
data is represented by the clustering. It does not consider the
quality with respect to the investment results. Based on the input
data considerations, they conclude that the k-means clustering using
median representative outperforms other clustering procedures,
independently of the number of clusters.

\change{In \cite{15}, different clustering based aggregation techniques each
selecting days are also compared}. Demand, solar and wind timeseries are considered in
the aggregation, each normalized such that they are equally important
in the aggregation. Three aggregation techniques are considered. A
hierarchical clustering with a minmax linkage criterion and a dynamic
time wrapping distance metric. A double clustering strategy first
applying a k-means clustering, followed by a re-clustering using the
aforementioned hierarchical procedure. The last technique is a pure k-means clustering. Each aggregated timeseries consists of 30 selected
days, and the comparison is based on test runs in a capacity expansion
model assuming continuous investment decisions and without unit
commitment decisions. The case study is based on a single dataset
covering three regions. The aggregation methods are compared on their
ability to replicate investments, which are possible in wind, solar,
coal, natural gas, nuclear and a generic storage technology. The mean
element is selected from each cluster, and selected elements are
weighted (repeated) according to the number of elements in the
clusters. They conclude that the hierarchical clustering and the
double clustering have similar performance, which is better than the
one obtained for a k-means clustering.

In \cite{20}, five different day selecting clustering approaches are
compared. First, a k-means and a k-medoids clustering each based on an
Euclidean distance metric and cluster centers as representative
elements. Then, a dynamic time warping barycenter averaging
clustering, a k-shape clustering with a shape-based distance metric
and lastly, a hierarchical clustering with Euclidean distance metric
and medoid element as cluster representative. Only price timeseries
are considered for the aggregation and each timeseries is normalized
(daily) using a z-normalization. Each clustering procedure is
considered for 1-9 clusters, but the quality of the approaches is
measured only by comparing the selected data to the original data,
i.e. in a data validation framework \cite{2}. After having decided on the
number of clusters, the techniques are compared based on their
resulting investments. The comparison is based on two different MIP
models representing different energy systems: one based on an
electricity storage and one based on a gas turbine power
generation. Furthermore, each aggregation is applied to each system
for two different timeseries of electricity prices. They conclude that
the centroid-based clusterings better replicate the operational part
while they also tend to underestimate objective values compared to the
medoid-based approaches.

In \cite{2}, three clustering procedures are compared to four
non-clustering aggregation techniques. The procedures are carefully
selected from the literature to represent classes of proposed
methods. Also, three new aggregation techniques are proposed; one
based on dynamically blocking days, one based on optimizing the
statistical representation of selected days; and one based on double
clustering including correlation as distance measure.  All methods are
compared on three test instances inspired by the Danish power system
and with different amounts of RES. The goal of the study is a broader
comparison of methods to provide general insight in the performance of
aggregation techniques. A conclusion from the study is that even
though a double clustering procedure is concluded to provide overall
best performance, the much simpler aggregation approaches show to be
strong competitors.

\subsection{\change{Hypothesis} and contribution of this paper}

\change{The common approach in the literature is to solve specific energy systems, which makes it difficult to compare the results. \cite{4} conclude that k-means has best performance, \cite{15} hierarchical and double clustering, in \cite{20} medoid based selection is best for investment decisions and \cite{2} show that simple heuristic based aggregation approaches perform well. In \cite{20} they do compare across different energy systems, but these energy systems are based on different underlying MIPs which again makes it difficult to draw conclusions for the CEP. }

\change{A similar pattern can be seen for element selection strategy. Comparisons of different
selection criteria is not common in the literature.} Exceptions are
\cite{14} considering median and mean representatives with median as best option, \cite{Nahmmacher2016} considering
selection of centroid or historical day representation with centroid as best option, and \cite{2}
considering minimum, maximum, mean and random element
selections with random as best option. 

Moreover, to our knowledge, the question of selecting
multiple elements from each cluster has not been addressed until the
current paper. In both \cite{14} and \cite{15}, the cluster
representative is weighted according to the cluster sizes, however
without argumentation nor analysis of why. To the best of our
understanding, weighting is performed by repeating each cluster
representative instead of selecting multiple elements from the
clusters. ~\\

\change{A different view on solving problems can be taken from the research field of optimizing solvers. Here, the goal is not to solve each problem to perfection, but to find an approach, which provides  overall good performance among different problems \cite{anand2017}. This is also the focus of this paper.}

The main contribution of this paper is a detailed and structured
comparison of different time aggregation approaches on four very
different energy systems, based on the same mathematical formulation. With the energy
systems being realistic in size and detail, the conclusions are widely
applicable. Since the aggregation technique comparison also includes a
very simple approach, this paper furthermore illuminates the relation
between aggregation technique complexity and performance. The paper
further contributes by comparing different selection strategies when
elements are to be selected from each cluster and by considering both
single and multiple selections from each cluster. Also, to the best of
our knowledge, this paper is the first to illustrate the benefits of
considering clustering weightings in the selection.

\change{
This paper focuses on the energy simulation model, Sifre. The full
mathematical formulation of Sifre is available at \cite{10} and the
capacity expansion module of Sifre is available in Appendix A.  Investment decisions are supported for production units, renewable units, storage, electric vehicles and
interconnection lines. }

\change{
When solving the investment problem, a full year is simulated. Sifre
LP relaxes the problem to limit the simulation solution time. The
integer variables in the unit commitment problem are LP relaxed and
investments are linear instead of discrete (e.g. invest between 0 and
500 MW in a production unit, instead of investing in zero, one or two
production units, each of size 150 MW). Still, solving the problem may
take many hours because of the problem instance size.}

\change{
This paper implements the aggregation techniques as part of Sifre, but
we still consider the results and analyses valid for other energy
system simulation models like e.g. the TIMES or Balmorel models \cite{4,7}.
}

\section{Solution methods}
\label{sol}
Numerous aggregation approaches are suggested in the
literature. Buchholz et al. \cite{2} survey the many approaches and
computationally compare aggregation strategies from the literature.
According to their studies, the following approaches show superior
performance:
\begin{itemize}
	\item \emph{Dummy Selection}, where every 13th element is selected from the residual load curve
	\item \emph{Statistical Representation}, which selects 10000 random samples and from this select the sample that best represents the means and standard deviation of the original data
	\item \emph{Optimized Selection}, which has same objective as Statistical Representation. Instead of investigating 10000 random samples, this approach finds the optimal sample with respect to means and standard deviation of the original data
	\item \emph{k-means Clustering} with squared Euclidean distance measurement
	\item \emph{Cluster Clustering} which first applies k-means
          clustering with squared Euclidean distances. Each resulting
          cluster is re-clustered using hierarchical agglomerative
          clustering with dynamic time warp distance measure and
          complete linkage criterion (minimizes the maximum distance
          between two elements; one in each cluster)
	\item \emph{Level Correlation Clustering}, which first applies fuzzy
          clustering with squared Euclidean distances. Then it applies
          hierarchical agglomerative clustering according to element
          correlations
\end{itemize}
To scope the work in this paper, we decide to focus on the clustering
methods (the three last methods). We also include Dummy Selection due
to its simplicity. As the sector coupled energy systems consist of
many timeseries (and not just the residual load), we select every 13th
element from each timeseries in Dummy Selection.

\subsection{Configuration of the clustering approaches}
\label{conf}
The survey of Buchholz et al. \cite{2} shows promising results when clustering days into 28 clusters. We thus apply this configuration. Both the
Cluster Clustering and the Level Correlation Clustering generates 7
outer clusters, each of which are re-clustered into 4 sub clusters.

The k-means clustering and fuzzy clustering algorithms depend on an
initial cluster. We divide the simulation period evenly into the
number of desired clusters. E.g. consider a year of 365 days, where
the number of desired clusters is 28 and where days are
clustered. Then the first 13 days are assigned to the first cluster,
the next 13 days to the next cluster etc.

\change{
\subsection{Data dimensions}
}
\change{The proposed methods are extended to handle complex sector coupled
energy systems, by making them consider all fluctuating timeseries data.} 
This means that the methods consider demand of all energy
types (e.g. power, district heating, gas), RES production,
import/export prices, fuel prices and availability profiles for
production units and interconnection lines. The clustering approaches
consider every fluctuating timeseries separately (instead of summing
them into e.g. a residual load curve) and we also maintain the
chronological order (in contrast to duration curves).  Demand is
negated to make the selection of cluster elements more intuitively
understandable. The minimum sum element in a cluster represents a day
with low production and high demand. Similarly, the maximum sum
element represents a day with high production and little
demand.  The clustering approaches must calculate the distance
between two days. This is done for each matching pair of timeseries
for each hour (e.g. the RES production by \change{offshore wind park} Horns Rev 1 for each
of the two days). All differences are summed across hours and
timeseries to produce the final distance between the two cluster days.

\subsection{Selecting days from clusters}
\label{select}
Two approaches can be considered for deciding the number of days to
select from each cluster. Either one day from each cluster (denoted
non-weighted or fixed weighted), or a weighted number of days from
each cluster. The benefit of the latter is that typical days and
outliers in the full dataset remain (somewhat) typical and outlying in
the aggregated dataset.  The weight is set according to the cluster
size: 
\begin{align}
\text{frequency} = \frac{\text{total number of days in simulation}}{\text{number of clusters}}\label{eq:freq}\\
 \text{weight} = \max \left(1, \text{round}\left(\frac{\text{cluster size}}{\text{frequency}}\right)\right)\label{eq:weight}
\end{align}

The total number of selected days may exceed the number of clusters for weighted selection. Weighted selection in the literature consists of selecting a single element from each cluster and then repeating this element a number of times \cite{Nahmmacher2016,14,15}. We propose to instead select a weighted number of elements from each cluster. The benefit of this is
that time chronology is maintained, i.e., once selection of elements
has finished, the original order of the selected elements is
applied. Also, selecting existing elements instead of generating new,
should represent the original data better.

Several strategies are investigated for deciding which days to select
from each cluster: Minimum sum, i.e. the day(s) with smallest sum;
Maximum sum, i.e. the day(s) with largest sum; Median sum, i.e. the
day(s) with median sum; Closest to Cluster Mean, i.e. the day(s) with
shortest distance to the cluster mean, and Random i.e. randomly chosen
day(s). Closest to Cluster Mean is calculated as follows: The mean of
a day is calculated for every hour. The distance from an element to
the mean is the total Euclidean distance in the 24-\change{dimensional} space.
\DP{Jeg har indføjet "dimensional". Er det korrekt forstået}

\change{
\subsection{Test setup} 
}
\label{compar}
\change{
The time aggregation techniques are compared to the optimal solution
of each data instance, i.e. we apply \emph{model validation} to measure the
quality of the aggregation techniques \cite{2}. Since only part of the
problem is solved by the time aggregation techniques, the objective
function values cannot be compared out of the box. It is, however, possible to
generate two full year simulations with fixed investments: one
simulation with optimal investments and another simulation with
investments from using a time aggregation technique. The objective
function values of these two simulations can then be compared. But
the objective function values will not include investment costs and
will thus be difficult to understand in relation to investment
decisions. Also, the objective function value is of very little
interest in the analyses in Energinet, where focus is on the energy mix, the
flows, etc. For this reason, we decide to only compare
the investment decisions. The performance measure hence becomes:
}
\change{
\begin{equation}
\label{comparisonMeasure}
\frac{\sum_i |x_i - \bar{x}_i|}{\sum_i \bar{x}_i}
\end{equation}
}

\change{
\noindent where $i$ is an index for the investments,  $x_i$ is the investment decision
made by the aggregation technique and $\bar{x}_i$ the investment decision from
the optimal solution.
}

\section{Test instances}
\label{test}
The aggregation techniques are tested on four significantly different sector coupled energy systems, all stemming from analyses in Energinet. The energy systems are different instances of the LP model summarized in Section \ref{cep}. An energy system consists of the following components:
\begin{itemize}
	\item \emph{Areas}, which represent an energy type and a geographical region, possibly attached an energy demand  
	\item \emph{External areas} represent an energy type and a geographical region. They only have a price per MWh for each hour attached. They can only be connected to the rest of the system via an interconnection line
	\item \emph{Production units} (or \emph{Conversion units} or \emph{Generation units}) convert energy types; examples are CHPs, CCGTs and compressors,
	\item \emph{Renewable units} produce energy based on a production profile
	\item \emph{Storages} are any types of storages, e.g., batteries or water tanks. Storages can also be used to model line pack in gas systems
	\item \emph{Electric vehicles} \change{which must be charged before requested driving time, however, the time of charging is flexible} \DP{Forklar lidt mere om hvordan elektriske biler adskiller sig fra normalt forbrug. Er de intelligente så de kan lades op på et vilkårligt tidspunkt? Kan de bruges som lager?} \MG{Er det bedre nu?}\DP{Jeg har uddybet. Er det korrekt forstået?}
\end{itemize}
The sector coupled energy systems are described in Table \ref{inst}.

\begin{table}[htbp]
    \small
    \centering
\caption{\label{inst}Statistics for test instances}
\begin{tabular}{l|c|c|c|c}\hline
Energy system &	Areas&	External areas&	Production units&	RES\\\hline
DK classic&	74&	6&	294&	60\\
DK detailed&	211&	9&	396&	88\\
Gas &	74&	2&	70&	7\\
PtX&	27&	8&	31&	1\\\hline
\multicolumn{5}{c}{}\\
\hline
&		Storages&	Interconnectors&Electric vehicles&	Demands\\\hline
DK classic&	36&8&	2&		66\\
DK detailed&	54&109&		16&	94\\
Gas&	6&11&		0&	15\\
PtX&	16&10&		0&	3\\\hline
\end{tabular}
\end{table}

\begin{figure}[htbp]
    \centering
    \includegraphics[scale=0.8]{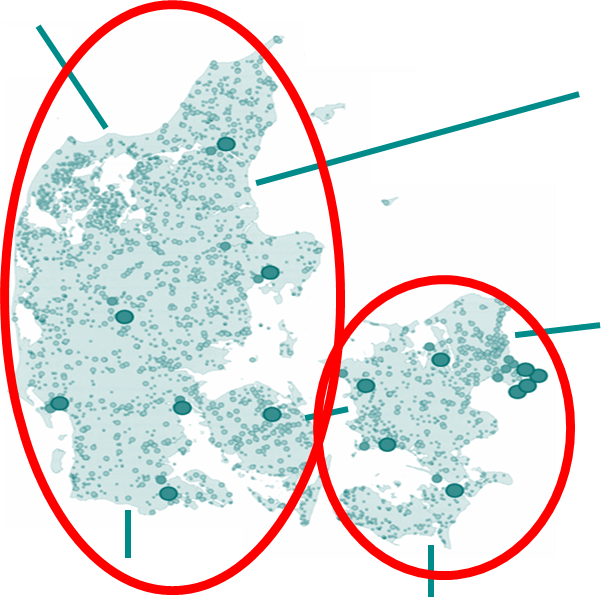}
    \caption{An overview of the DK classic instance. The red circles
represent electricity areas, the blue lines interconnection lines to
neighboring electricity areas. The blue dots show district heating
areas in Denmark, which are modelled as 59 district heating areas in
the dataset.}
    \label{fig1}
\end{figure}

The \emph{DK classic} instance consists of a representation of the Danish
power and district heating system in 2020, see Figure \ref{fig1}. The
investment decisions focus on heat production and consist of two CHPs,
three heat boilers and three heat pumps: a total of 8 investments.

\begin{figure}[htbp]
    \centering
    \includegraphics[scale=0.8]{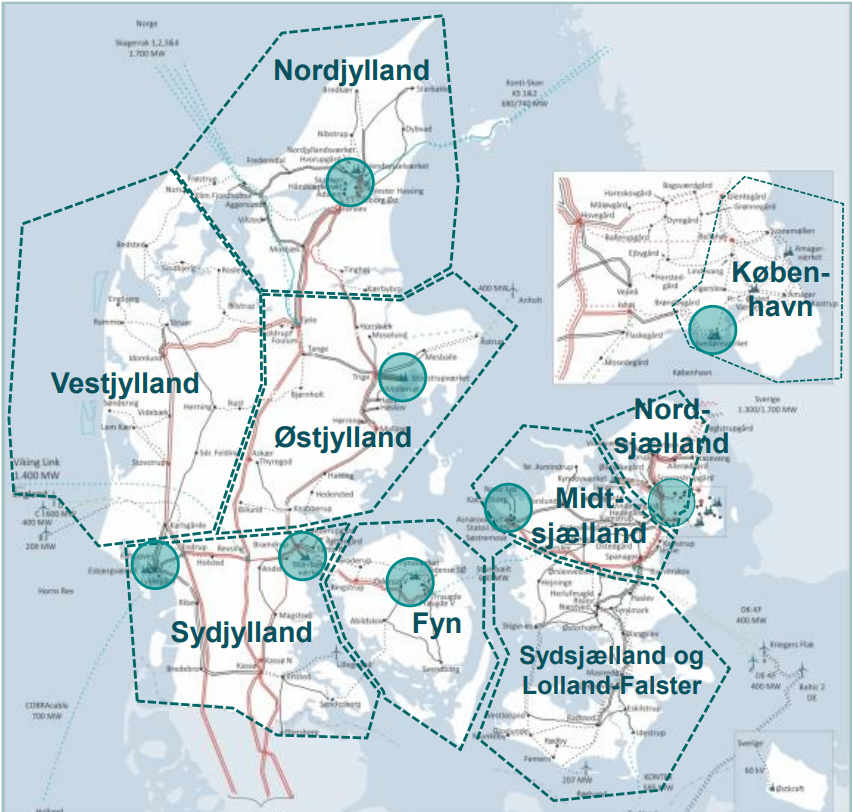}
    \caption{An overview of the DK detailed instance. The electricity
areas are highlighted. District heating is modelled as 59 areas, as for DK Classic.}
    \label{fig2}
\end{figure}

The \emph{DK detailed} instance consists of the Danish power and district heating
system in 2050, see Figure \ref{fig2}. The number of electricity areas are
split into eight areas to represent possible future grid
bottlenecks. Also, the production system includes PtX technologies
(Power to X technologies), hence fuels are represented in greater
detail than in DK classic and include parts of the transportation
sector. The investment decisions focus on seven PtX plants, modelled
through fourteen condensing power plants, seven heat pumps and one
storage: a total of 22 investments.

\begin{figure}[htbp]
    \centering
    \includegraphics[scale=0.35]{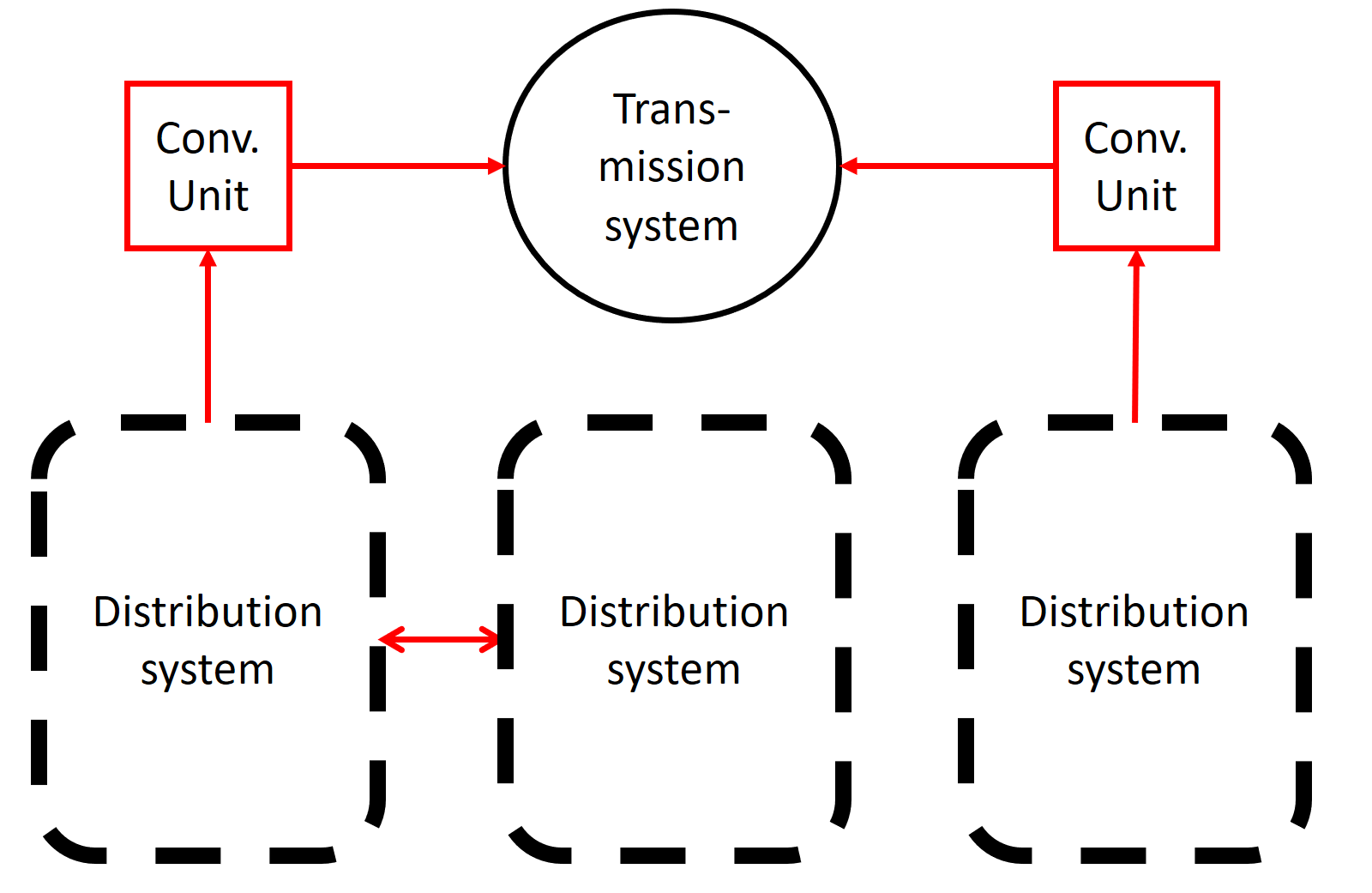}
    ~\\~\\
    \includegraphics[scale=0.3]{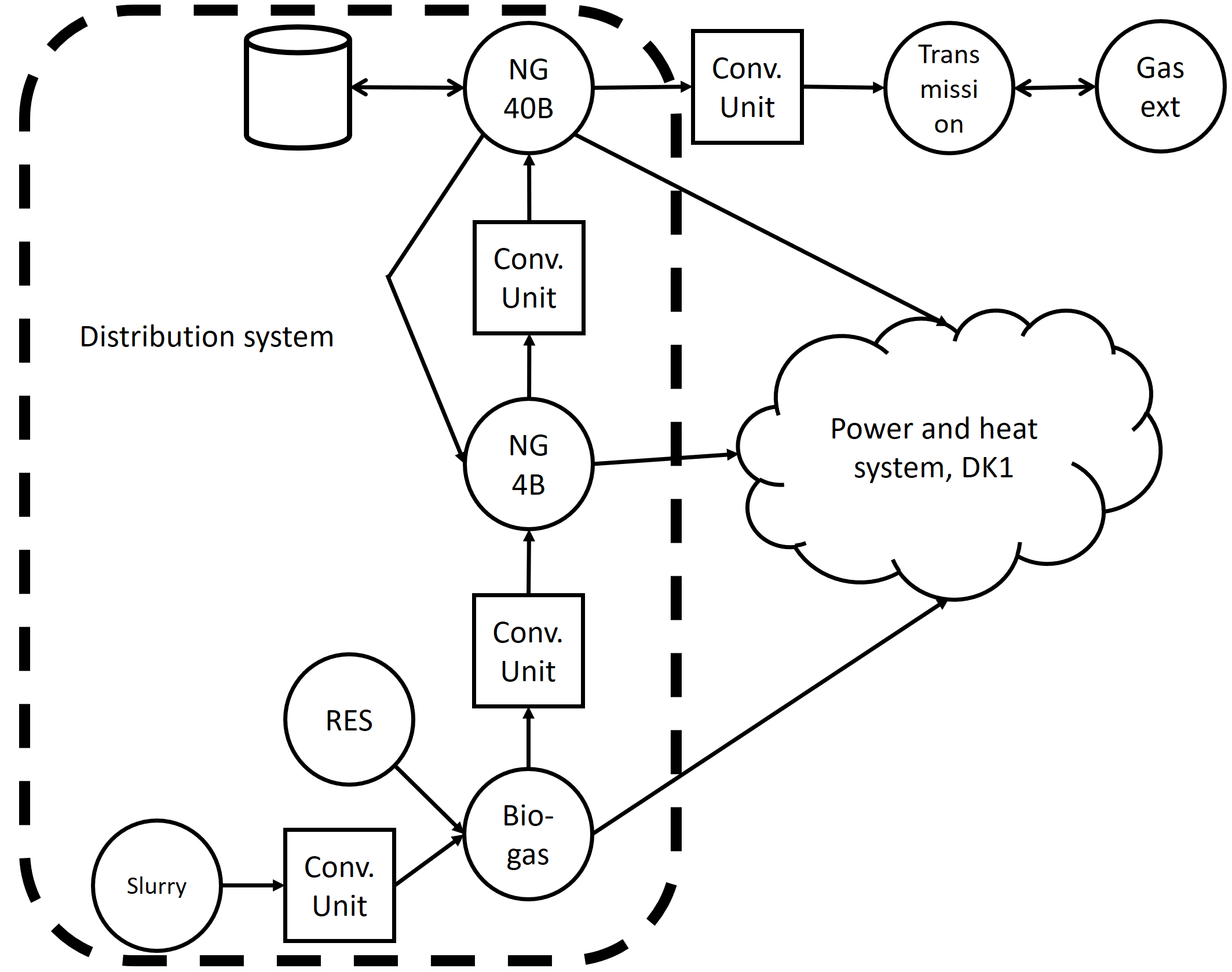}
    \caption{An overview of the Gas instance. The upper figure illustrates
the overall system. A total of 10 distribution systems are modelled in
varying detail. The investments are colored red. The lower figure
illustrates an example of a modelled gas distribution system, where NG
is short for natural gas, 40B is 40 bar and 4B is 4 bar.}
    \label{fig3}
\end{figure}

The \emph{Gas} instance consists of a subpart of the Danish gas transmission
and distribution systems in 2020, see Figure \ref{fig3}. The instance
introduces large amounts of biogas and investigates investments in two
compressors from gas distribution systems to the gas transmission
system and one investment in connecting distribution systems directly:
a total of 3 investments.

\begin{figure}[htbp]
    \centering
    \includegraphics[width=\columnwidth]{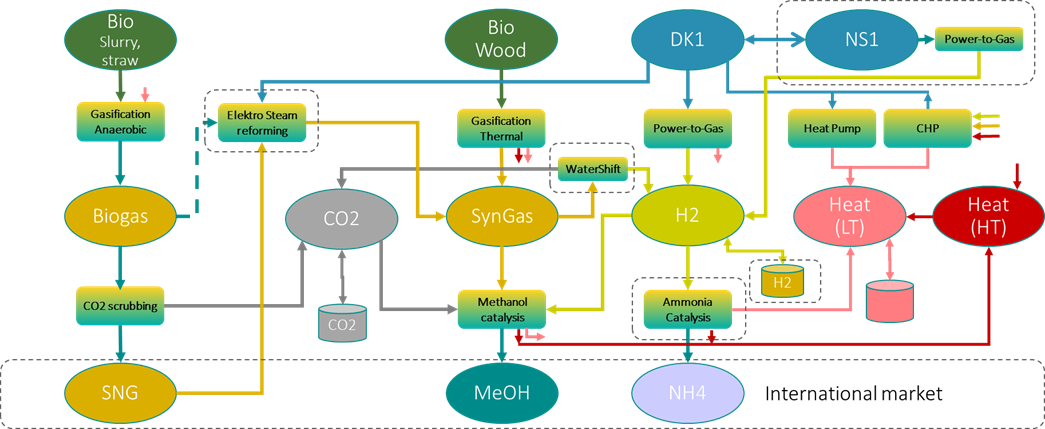}
    \caption{An overview of the PtX instance. The system integrates many
energy types.}
    \label{fig4}
\end{figure}

The \emph{PtX} instance models a Power to X cluster as illustrated in Figure \ref{fig4}. The investment
possibilities decide how to dimension the PtX cluster and consists of
19 production units, one heat pump and one interconnection line: a
total of 21 investments.

\section{Results}\label{res}

The computational evaluation is conducted on a 10 core 2,4 GHz machine
with 128 GB RAM, using Gurobi 8.1 as solver. The following
abbreviations are used in the remainder of this section: 
\emph{Clustering methods:}
\textbf{k} for k-means, \textbf{cc} for Cluster Clustering, and \textbf{lc} for Level
Correlation clustering. 
\emph{Selection strategies:}
\textbf{min} for minimum sum, \textbf{max}
for maximum sum, \textbf{median} for median sum, and \textbf{cmean} for closest to
cluster mean. Finally, we have \textbf{w} for weighted selection, \textbf{n} for
non-weighted (fixed weighted) selection, and \textbf{28} to represent the 28
generated clusters. Run time results are seen in Table \ref{t1} and solution quality gaps in Table
\ref{t2}. Results are analyzed in the following sections.

\begin{table}[htbp]
    \small
\centering
\caption{\label{t1}Run times in minutes.}
\begin{tabular}{l|rrrr}
 	&DK-classic	&DK-detailed	&Gas	&PtX\\\hline\hline
Full test instance	&36,45	&536,06	&11,71	&14,69\\\hline
Dummy Selection	&4,38&	18,29&	1,48&	1,47\\\hline
k,min,w,28&	4,21&	17,42&	1,34&	1,54\\
k,max,w,28&	5,83&	22,65&	1,32&	1,45\\
k,median,w,28&	3,99&	24,64&	1,38&	1,49\\
k,cmean,w,28&	4,25&	15,42&	1,46&	1,65\\
k,random,w,28&	4,10&	22,32&	1,39&	1,45\\\hline
cc,min,w,28&	5,74&	43,48&	1,63&	1,07\\
cc,max,w,28&	5,58&	25,04&	1,66&	1,49\\
cc,median,w,28&	6,16&	48,25&	1,62&	1,06\\
cc,cmean,w,28&	6,04&	33,26&	1,67&	1,51\\
cc,random,w,28&	5,66&	31,73&	1,71&	1,21\\\hline
lc,min,w,28&	6,21&	35,76&	1,54&	2,15\\
lc,max,w,28&	8,20&	27,42&	1,93&	1,53\\
lc,median,w,28&	7,01&	21,86&	1,90&	1,68\\
lc,cmean,w,28&	6,70&	35,30&	1,90&	1,07\\
lc,random,w,28&	7,34&	34,78&	1,91&	1,82\\\hline
k,min,n,28&	    5,95&	10,19&	1,26&	1,27\\
k,max,n,28& 	3,89&	16,82&	1,25&	0,88\\
k,median,n,28&	5,35&	16,72&	1,24&	0,86\\
k,cmean,n,28&	5,72&	17,40&	1,25&	0,86\\
k,random,n,28&	4,21&	15,35&	1,26&	1,17\\\hline
cc,min,n,28&	6,56&	21,20&	1,36&	1,15\\
cc,max,n,28&	6,48&	19,24&	1,44&	1,01\\
cc,median,n,28&	5,08&	18,80&	1,45&	0,86\\
cc,cmean,n,28&	5,19&	15,89&	1,41&	0,76\\
cc,random,n,28&	5,24&	25,55&	1,40&	0,77\\\hline
lc,min,n,28&	7,01&	21,67&	1,71&	1,14\\
lc,max,n,28&	4,83&	17,72&	1,75&	1,05\\
lc,median,n,28&	6,26&	24,05&	1,72&	1,11\\
lc,cmean,n,28&	6,41&	35,26&	1,74&	0,93\\
lc,random,n,28&	6,54&	21,10&	1,74&	1,00\\\hline
\end{tabular}
\end{table}

\begin{table}[htbp]
    \small
\centering
\caption{\label{t2}Solution quality: the lower percentage, the better performance.}
\begin{tabular}{l|rrrr}
 	&DK-classic	&DK-detailed	&Gas	&PtX\\\hline\hline
Dummy Selection&	8\%&	5\%&	25\%&	4\%\\\hline
k,min,w,28&	    101\%&	3\%&	47\%&	2\%\\
k,max,w,28&	    65\%&	18\%&	25\%&	3\%\\
k,median,w,28&	17\%&	4\%&	24\%&	0\%\\
k,cmean,w,28&	19\%&	6\%&	39\%&	2\%\\
k,random,w,28&	16\%&	8\%&	40\%&	2\%\\\hline
cc,min,w,28&	105\%&	4\%&	38\%&	1\%\\
cc,max,w,28&	30\%&	23\%&	12\%&	7\%\\
cc,median,w,28&	17\%&	2\%&	37\%&	2\%\\
cc,cmean,w,28&	13\%&	5\%&	35\%&	1\%\\
cc,random,w,28&	17\%&	5\%&	28\%&	2\%\\\hline
lc,min,w,28&	23\%&	4\%&	61\%&	1\%\\
lc,max,w,28&	90\%&	23\%&	17\%&	5\%\\
lc,median,w,28&	31\%&	6\%&	32\%&	1\%\\
lc,cmean,w,28&	42\%&	9\%&	28\%&	0\%\\
lc,random,w,28&	6\%&	2\%&	23\%&	0\%\\\hline
k,min,n,28& 	102\%&	4\%&	51\%&	1\%\\
k,max,n,28&	    66\%&	7\%&	40\%&	6\%\\
k,median,n,28&	9\%&	11\%&	23\%&	1\%\\
k,cmean,n,28&	13\%&	4\%&	41\%&	3\%\\
k,random,n,28&	5\%&	2\%&	25\%&	3\%\\\hline
cc,min,n,28&	75\%&	7\%&	51\%&	2\%\\
cc,max,n,28&	74\%&	17\%&	49\%&	9\%\\
cc,median,n,28&	45\%&	2\%&	59\%&	4\%\\
cc,cmean,n,28&	23\%&	3\%&	57\%&	4\%\\
cc,random,n,28&	37\%&	6\%&	55\%&	8\%\\\hline
lc,min,n,28&	33\%&	4\%&	68\%&	1\%\\
lc,max,n,28&	57\%&	20\%&	49\%&	66\%\\
lc,median,n,28&	40\%&	6\%&	35\%&	2\%\\
lc,cmean,n,28&	47\%&	8\%&	29\%&	2\%\\
lc,random,n,28&	49\%&	7\%&	33\%&	3\%\\\hline
\end{tabular}
\end{table}

\subsection{Time usage}
\begin{figure}[htbp]
    \centering
    \includegraphics[width=0.9\columnwidth]{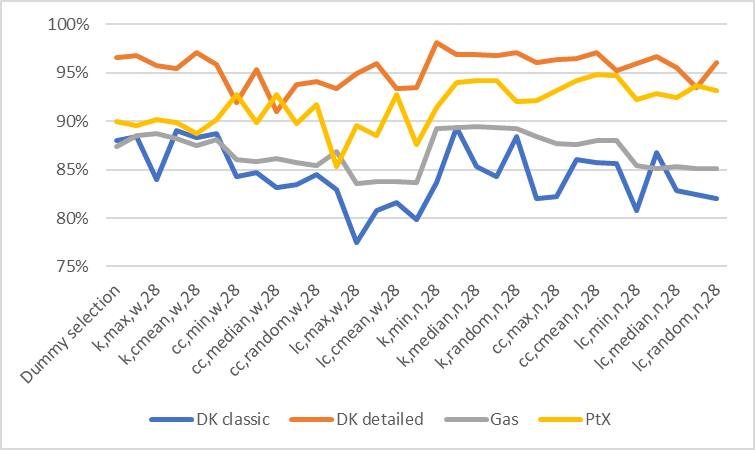}
    \caption{\label{fig5}Time usage reductions in percent.}
\end{figure}

Time reductions are plotted
in Figure \ref{fig5}. Note that the solution times also include pre- and
postprocessing of the data instances and not only time for solving the linear
program. The time usage savings are consistent across the time
aggregation techniques. The average time saving is 90\%, which is very
satisfying. The time savings are slightly smaller for the DK classic
and Gas instances, which could indicate that these instances spend
relatively more time on pre- and postprocessing data than the DK
detailed and PtX instances.

Generally, the time
savings are slightly smaller for the weighted selection
algorithms. Recall the weighting from Section \ref{select}; rounding the number
of elements to select from a cluster may increase the total number of
selected days. Indeed, the weighted selections result in more than 28
selected days, see Table \ref{t3}.
\begin{table}[htbp]
    \small
\centering
\caption{\label{t3}The number of selected days.}
\begin{tabular}{l|rrrr}
 	&DK classic&	DK detailed&	Gas&	PtX\\\hline\hline
Dummy selection&	28	&28	&28	&28\\
k,n,28&         	28	&28	&28	&28\\
cc,n,28&        	28	&28	&28	&28\\
lc,n,28&        	28	&28	&28	&28\\\hline
k,w,28&         	31	&34	&34	&36\\
cc,w,28&        	37	&38	&39	&39\\
lc,w,28&        	38	&37	&36	&38\\\hline
\end{tabular}
\end{table}

\subsection{Weighted vs. non-weighted selection}
\begin{figure}[htbp]
    \centering
    \includegraphics[width=0.9\columnwidth]{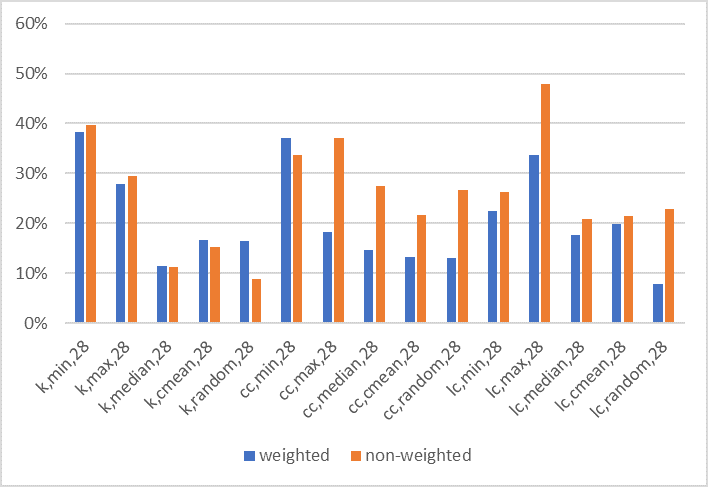}
    \caption{\label{fig6}Quality gap percentages averaged across the four instances.}
\end{figure}

Weighted selection has better performance than non-weighted selection
with respect to solution quality in 62\% of the time aggregated
simulations. The results are illustrated in Figure \ref{fig6}. In 37 of 60
cases, the investment gap decreases with weighted selection. If gaps
are averaged across instances, the gap decreases with weighted
selection in 11 out of 15 cases. The average of all gaps is 21\% for
weighted selection and 26\% for non-weighted selection. This confirms
that weighted selection better represents the full dataset and that
outliers are balanced well against the rest of the dataset. The
improved quality may partly be due to the increased number of selected
days, see Table \ref{t3}. It is possible to increase the number of selected
days for the non-weighted algorithms and compare the results. This
would, however, require that the non-weighted algorithms generate more
clusters, which again would make comparison more difficult. Instead we
continue to compare the algorithms with 28 clusters. The interested
reader is referred to Appendix \ref{app_comp} for results for non-weighted
selection with more clusters.

\subsection{Selection strategy}
\begin{figure}[htbp]
    \centering
    \includegraphics[width=0.9\columnwidth]{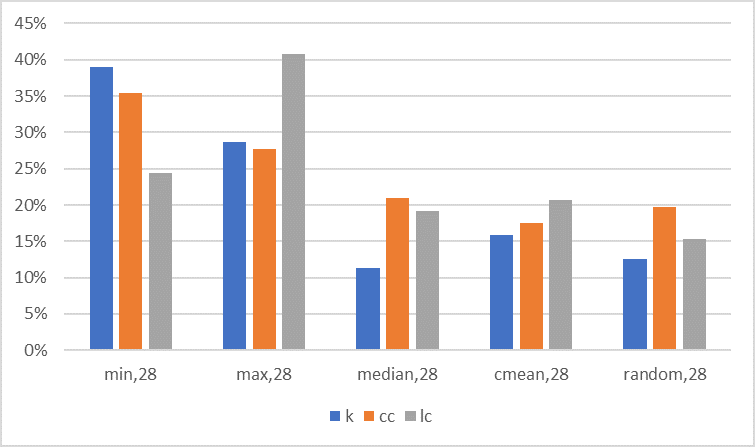}
    \caption{\label{fig7}Quality gap percentages averaged across instances and weighted and non-weighted selection.}
\end{figure}

\begin{figure}[htbp]
    \centering
    \includegraphics[width=0.9\columnwidth]{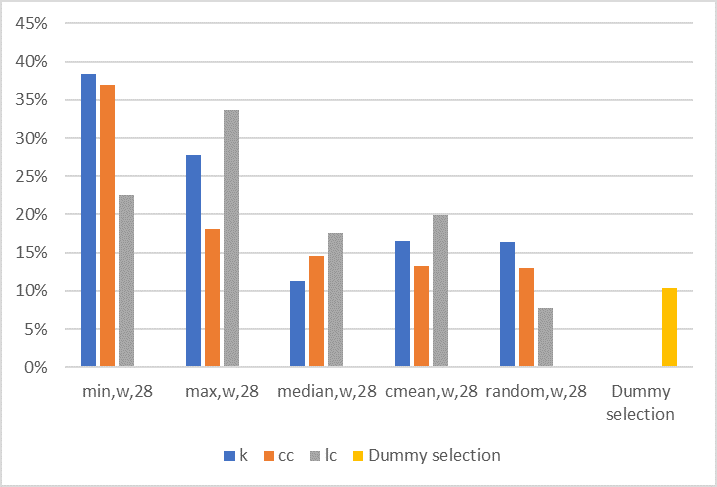}
    \caption{\label{fig8}Quality gap percentages averaged across instances for weighted selection only.}
\end{figure}

The strategies for selecting elements in each cluster perform
differently across the instances. Results averaged across the four instances are illustrated in Figure \ref{fig7}. Clearly, the minimum sum and maximum sum selections have worst
performance. Random and median selection vary slightly, while closest
to cluster mean gives consistent results. The same pattern is seen, when considering results for weighted
selection only, see Figure \ref{fig8}. Selecting only the minimum or maximum
sum elements represents the clusters less well. Random performs well
which indicates that always selecting the median or closest to cluster
mean elements may be too strict.

\subsection{New approaches to promote diversification in selected days}
\label{new}
Random selection performs well but due to its random nature, results
are not consistently good. To eliminate the randomness, we instead
seek to mimic differentiated selection.

We propose the MedianMaxMin selection. The approach is only relevant
in weighted selection where more than one element may be selected from
each cluster. First the median element is selected. If more elements
are to be selected from the cluster, the maximum element is
selected. Again, if more elements are to be selected, the minimum
element is selected. If even more elements are to be selected from the
cluster, the selection order repeats.

We also propose the kk-means clustering approach (in short kk). The
outer clustering is k-means with squared \change{Euclidean} distances where the
initial clusters are generated as explained in Section \ref{conf}. The inner
clustering is also a k-means with squared Euclidean distances, but
this time the initial clusters are formed around the median, maximum
sum and minimum sum elements (in the outer cluster).

\begin{figure}[htbp]
    \centering
    \includegraphics[width=0.9\columnwidth]{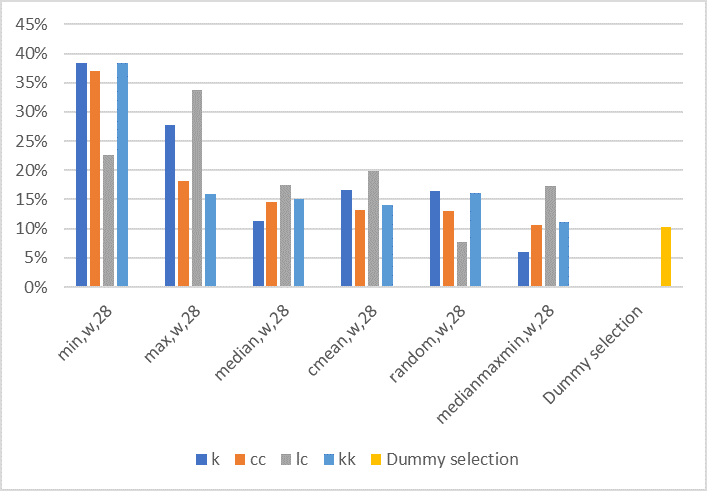}
    \caption{\label{fig9}Quality gap percentages averaged across instances for weighted selection and dummy selection.}
\end{figure}

The two approaches are tested. 
Run times are seen in Table \ref{t6} and
solution gaps in Table \ref{t7}. Run times are consistent with the remaining aggregation approaches.  \change{Solution gaps
are illustrated in Figure \ref{fig9}}. MedianMaxMin selection
generally performs better than the other selections strategies. Good
results are especially achieved together with k-means, cluster
clustering and kk-means. \change{Kk-means performs overall well, however, }without
outperforming the other clustering approaches. It gives consistent
results except for min selection, which generally performs poorly
regardless of clustering approach.

\begin{table}[htbp]
    \small
\centering
\caption{\label{t6}Run times in minutes.}
\begin{tabular}{l|rrrr}
 	&DK-classic	&DK-detailed&	Gas&	PtX\\\hline\hline
k,MedianMaxMin,w,28&	5,67&	28,39&	1,30&	1,52\\
cc,MedianMaxMin,w,28&	5,45&	21,55&	1,65&	2,26\\
lc,MedianMaxMin,w,28&	5,27&	27,24&	1,57&	1,28\\\hline
kk,min,w,28&        	3,86&   23,35&	1,48&	1,20\\
kk,max,w,28&        	7,35&   21,78&	1,48&	1,36\\
kk,median,w,28&     	5,56&	13,88&	1,50&	1,45\\
kk,cmean,w,28&      	5,56&	26,14&	1,61&	1,70\\
kk,random,w,28&     	4,07&	24,19&	1,54&	1,26\\
kk,MedianMaxMin,w,28&	6,33&	23,10&	1,51&	1,35\\\hline
\end{tabular}
\end{table}

\begin{table}[htbp]
    \small
\centering
\caption{\label{t7}Solution quality: the lower percentage, the better performance.}
\begin{tabular}{l|rrrr}
 	&DK-classic	&DK-detailed	&Gas	&PtX\\\hline\hline
k,medianmaxmin,w,28&	4\%&	4\%&	15\%&	1\%\\
cc,medianmaxmin,w,28&	16\%&	2\%&	21\%&	2\%\\
lc,medianmaxmin,w,28&	29\%&	7\%&	31\%&	1\%\\\hline
kk,min,w,28&        	104\%&	3\%&	43\%&	3\%\\
kk,max,w,28&        	41\%&	6\%&	7\%	&   9\%\\
kk,median,w,28&     	29\%&	9\%&	20\%&	2\%\\
kk,cmean,w,28&      	14\%&	2\%&	37\%&	3\%\\
kk,random,w,28&     	12\%&	2\%&	48\%&	3\%\\
kk,medianmaxmin,w,28&	7\%&	9\%&	23\%&	6\%\\\hline
\end{tabular}
\end{table}

\section{Further analysis}\label{fur}

\begin{figure}[htbp]
    \centering
    \includegraphics[width=0.7\columnwidth]{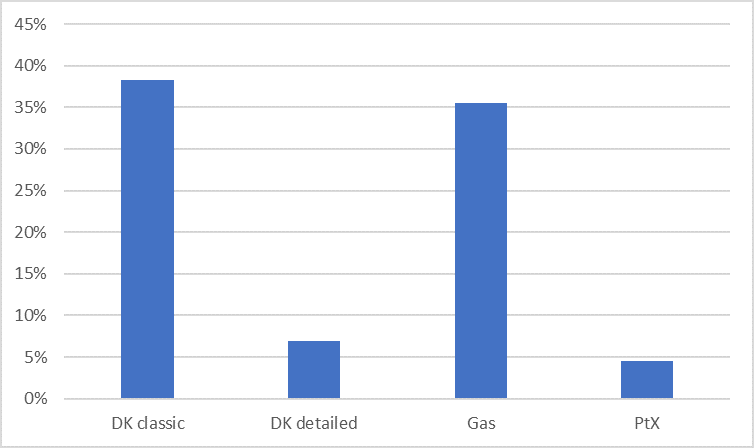}
    \caption{\label{fig10}Average solution gap for each instance across the clustering methods.}
\end{figure}

The results reveal different quality across time aggregation
techniques and data instance. The data instances are analyzed to
better understand the results; specifically, we analyze the behavior
of the investments in the optimal solution for the full test instances
and what this means to the clustering approaches. Average gaps for the
instances are seen in Figure \ref{fig10}.
\smallskip

\textbf{DK-classic:} The investments are mainly utilized in the winter
period. They are driven by district heating demand. Data in the
instance, however, also contains many other fluctuating timeseries,
especially connected to the electricity system: demand, RES
production, electricity prices in neighboring countries and capacity
restrictions on interconnection lines. The clustering methods end up
\change{generating clusters and} selecting days, which are not relevant to the investments.
\smallskip
	
\textbf{Gas:} The investments are utilized in the summer period, where demand
is low. In these hours, excessive biogas is either moved between
distribution systems or sold to the transmission system. Gas demand
is, however, not the only varying data in the instance. Electricity
prices (considered by gas-fueled CHPs) vary throughout the year. Line
pack is modelled as storage space with highest value in the
spring. Gas demand varies more outside the summer period (the higher
demand, the higher absolute variation). Hence the clustering
approaches end up \change{generating clusters and} selecting days from other seasons than the
summer and the solution quality suffers.
\smallskip

\textbf{DK-detailed:} The investments follow RES production and electricity
demand. Most timeseries in the dataset are related to RES production
and electricity, which explains the good solution quality.

\textbf{PtX:} The
fluctuation of the timeseries correspond well to the entire production
system, including the optimal investments. All time aggregation
methods thus have good performance.

\change{\subsection{Overall best aggregation method}}
\change{Given the analysis in this section and the results in the previous, we investigate which aggregation method show most promising results.}

\change{Dummy selection performs well but the method is not robust towards
investments, which are utilized in only part(s) of the simulated
year. This is the case for the Gas instance where Dummy Selection ends
up with a 25\% gap. For this reason, it may not generally be the best approach. It, however, benefits from being very simple to implement and to understand from the analyst's point of view.}

\change{For the clustering approaches, we have already concluded that best performance is achieve with weighted selection and with other selection methods than min and max. The overall best performing method in our survey is k-means clustering with weighted selection and with the MedianMinMax selection strategy. The k-means is simple to implement and  the MedianMinMax strategy diversifies selection without introducing the uncertainty of randomness. }

\change{We recommend this method but are also aware that this is a close call. This could indicate that the performance bottleneck no longer lies in the clustering or selection itself. This is investigated in the next section on future work.
}

\change{\subsection{Future Work}}
The clustering methods suffer from generating clusters based on data
fluctuations irrelevant to the investment decision. \change{We have identified three ideas to further dive into this.}

Future work could
focus on methods to better represent data. One method could be to
normalize data to take on values between e.g. -1 and 1. This could
lead to a fairer comparison of data stemming from different sources,
e.g. comparing capacities with prices. This would, however, also erase
the absolute amounts and thus treat e.g. large demands equally to
small demands. Fluctuations in small timeseries may cause unimportant
days to be selected and thus negatively affect the clustering
approach.  

Future work could also focus on dimensionality reduction,
e.g. by considering the subset of data needed to represent the
statistical behavior of each day, or by considering the subset of data
which correlates with the investment decisions.  

Finally, future work
could focus on testing the methods on instances with more
investment decisions, especially where the utilization of the
investment decisions is not correlated. This could possibly better
test the difference between the clustering methods, especially if
level correlation clustering performs differently than the other three
clustering methods.

\section{Conclusion}
\label{conc}
\change{In this paper, we have investigated the performance of clustering techniques across very different energy systems to give a recommendation of a method with overall good performance.}
We have applied a number of clustering techniques to real-life capacity expansion problems. The clustering techniques all select a subset of days from the datasets, which cover a full year. The applied methods are k-means, hierarchical clustering and a double clustering procedure
applying a fuzzy clustering, followed by a hierarchical clustering
considering element correlations. Also, we proposed a new method consisting of double k-means clustering. 

The methods cluster days and then selects a number of days from each cluster. We have tested several selection strategies from the literature: min, max, median, closest to cluster mean and random. We have also proposed a new selection strategy, MedianMaxMin, which selects elements in the named order. Finally, we have investigated the effect of selecting a single element from each cluster or a weighted number of elements from each cluster.

\change{All in all, this resulted in a comparison of 41 aggregation techniques, and the results were benchmarked against the full datasets.}
\MG{jeg tror ikke, at jeg forstår sætningen nu.}
\DP{Jeg har forsøgt at omformulere. Jeg var ikke så glad for "with each other" fordi det ligger implicit i ordet "comparison" at man sammenligner med hinanden}
The comparison is \change{evaluated} on how well
the investment decisions are matched. The methods were tested on four
very different energy systems to investigate performance consistency
and to analyze if certain energy system aspects are more difficult to
replicate through aggregation.

The tests showed that all aggregation techniques resulted in
significant time reductions between 78\% and 97\%. The tests also
revealed that weighted selection outperformed selecting exactly one
element from each cluster. To the best of our knowledge, this has not
been analyzed or concluded previously in the literature. Selecting
minimum or maximum elements from each cluster was generally not a good
strategy. The new selection method, MedianMaxMin, and clustering
method, kk, both performed consistently well. Especially k-means with
MedianMaxMin selection showed very good performance, \change{ and this is also the clustering approach we recommend.}

We also tested Dummy Selection, which simply selects every 13th day. It overall performed surprisingly well. Considering its simplicity, it could
be a good alternative to the more complex clustering methods as it is
easy to implement and understand.

Future work could focus on how data is considered when clustering. In
this paper, all timeseries are considered. A closer analysis of the
test instances revealed that this may not be the best approach as data
irrelevant to the investments caused the aggregation techniques to
select days, which were also irrelevant to the investment decisions.

\appendix

\newpage
\section{Capacity Expansion Modelling in Sifre}
This Appendix explains how capacity expansion is added to the energy
system model formulation in Sifre \cite{10}. The notation is kept at a high
level, such that the reader can understand the formulations without
reading the extensive nomenclature in \cite{10}. The set T defines the time
steps, and index i is used to denote the relevant component. Also, z
denotes the objective function for the energy system model formulation
in \cite{10}.
\medskip
	
\noindent\textbf{Storage}\\
The variable $\texttt{storage}_i$ decides the invested inventory level of storage
$i\in I$. The formulation adds the investment costs to the objective. The
constraints ensure that the invested amount satisfies the specified
lower and upper bound, and that the storage inventory level never
exceeds the invested inventory level.

\begin{figure}[htbp]
    \centering
    \includegraphics[scale=0.5]{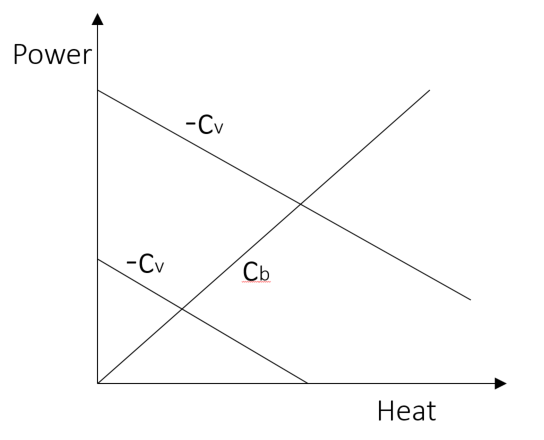}
    \caption{ PQ diagram to define the relationship between power and heat
production on a CHP. The operating area is bounded by the y-axis to
the left, the upper and lower $c_v$ lines and the $c_b$ line to the
right. The $c_v$ and $c_b$ are input parameters derived from possible
operations points of the CHP.}
    \label{figpq}
\end{figure}

\change{
\begin{align}
     \min \quad& z + \sum_{i\in I}\texttt{investmentCost}_i \cdot \texttt{storage}_i  \\
     \text{s. t.} \quad& lb_i \leq \texttt{storage}_i \leq ub_i&\forall i\in I \\
     & \texttt{invLevelVar}_i^t \leq \texttt{storage}_i& \forall i\in I, \forall t \in T
\end{align}
}
\noindent
\textbf{Interconnection line}\\
The variable $\texttt{icl}_i$ decides the invested capacity of the
interconnection line $i\in I$. The formulation adds the investment costs to
the objective function. The constraints ensure that the invested
amount satisfies the specified lower and upper bound, and that the
flow on the interconnection line never exceeds the invested capacity.
\begin{align}
     \min \quad& z + \sum_{i\in I}\texttt{investmentCost}_i \cdot \texttt{icl}_i  \\
     \text{s. t.} \quad& lb_i \leq \texttt{icl}_i \leq ub_i & \forall i\in I\\
     & \texttt{flowVar}_i^t \leq \texttt{icl}_i& \forall i\in I, \forall t \in T
\end{align}
\noindent
\textbf{Renewable units}\\
The variable $\texttt{res}_i$ decides the invested capacity of the renewable unit
$i\in I$. The formulation adds the investment costs to the objective
function. The constraints ensure that the invested amount satisfies
the specified lower and upper bound, and that the renewable production
never exceeds the invested capacity. The parameter $\texttt{profile}_i^t\in [0,1]$
defines the possible RES production subject to e.g. available wind or
solar. If curtailment is allowed, production in an hour does not have
to equal the available RES production.
\begin{align}     
\min \quad& z + \sum_{i\in I}\texttt{investmentCost}_i \cdot \texttt{res}_i  \\
     \text{s. t.}\quad& lb_i \leq \texttt{res}_i \leq ub_i&\forall i\in I \\
\omit\rlap{\text{if curtailment is allowed:}}  \nonumber   \\
     & \texttt{prodVar}_i^t \leq \texttt{profile}_i^t \cdot res_i& \forall i\in I, \forall t \in T\\
\omit\rlap{\text{if curtailment is not allowed:}}   \nonumber  \\
     & \texttt{prodVar}_i^t = \texttt{profile}_i^t \cdot res_i& \forall i\in I, \forall t \in T
\end{align}
\noindent
\textbf{Condensation plants, heat boilers, heat pumps, electric boilers and backpressure CHPs}\\
The following formulation is valid for production units, which either
only produces one type of energy, or where the relationship between
the primary and second energy production is fixed (e.g. power and heat
production in backpressure units). In the latter case, the investment
bounds are specified according to the primary production.  The
variable $\texttt{prod}_i$ decides the invested capacity of the production
unit $i$. The formulation in the following adds the investment costs to
the objective function. The constraints ensure that the invested
amount satisfies the specified lower and upper bound, and that the
production of the unit never exceeds the invested capacity.
\begin{align}
     \min \quad& z + \sum_{i\in I }\texttt{nvestmentCost}_i \cdot \texttt{prodCap}_i  \\
     \text{s. t.} \quad& lb_i \leq \texttt{prodCap}_i \leq ub_i &\forall i\in I \\
     & \texttt{prodVar}_{i,\text{primary}}^t \leq \texttt{prodCap}_i& \forall i\in I, \forall t \in T
\end{align}
\noindent
\textbf{Extraction CHPs}\\
The following formulation is valid for extraction production
units. The relationship between the primary and secondary energy
production (power and heat) is defined by a PQ diagram, see Figure \ref{figpq}.
 
The variable $\texttt{prodCap}_i$ decides the
invested capacity of the production unit $i \in I$. The formulation in the
following adds the investment costs to the objective function. The
constraints ensure that the invested amount satisfies the specified
lower and upper bound, and that the production of the unit never
exceeds the invested capacity. Both the primary and secondary
production must be considered to ensure the correct capacity for the
extraction unit.

\begin{align}
     \min \quad& z + \sum_{i \in I } \texttt{investmentCost}_i \cdot \texttt{prodCap}_i  \\
     \text{s. t.}\quad& lb_i \leq \texttt{prodCap}_i \leq ub_i & \forall i \in I\\
     & \texttt{prodVar}_{i,\text{primary}}^t + c_v \cdot \texttt{prodVar}_{i,\text{secondary}}^t \leq \texttt{prodCap}_i& \forall i \in I, \forall t \in T
\end{align}
\noindent where $c_v$ defines the upper bound relationship between heat and power production in the PQ diagram, see Figure \ref{figpq}.

\section{Results for non-weighted selection with more clusters} \label{app_comp}
Weighted selection algorithms end up selecting more elements than
non-weighted, see Section \ref{select}. The same number of selected elements
can be achieved by allowing the algorithms with non-weighted selection
to generate more clusters. For example, if an algorithm with weighted
selection ends up selecting 32 days, then we can configure the
algorithm with non-weighted selection to generate 32 clusters. As the
latter selects one element from each cluster, it ends up selecting 32
days.  The double clustering approaches sets the number of outer
clusters to the total number of desired clusters divided by 4. In case
of a total of 32 clusters, the approaches generate 8 outer
clusters. In case of a total of 29 clusters, the approaches generate 7
outer clusters. Each outer cluster is re-clustered into four inner
clusters. For this reason, the final number of clusters may variate
slightly from the desired amount for the double clustering
approaches. Furthermore, the total number of selected elements may
also vary slightly for the cc and kk approaches because the outer K
means algorithm may generate clusters with few elements, e.g. 1
element, and the inner clustering approaches can thus not generate
“enough” clusters.

\begin{figure}[htbp]
    \centering
    \includegraphics[width=0.9\columnwidth]{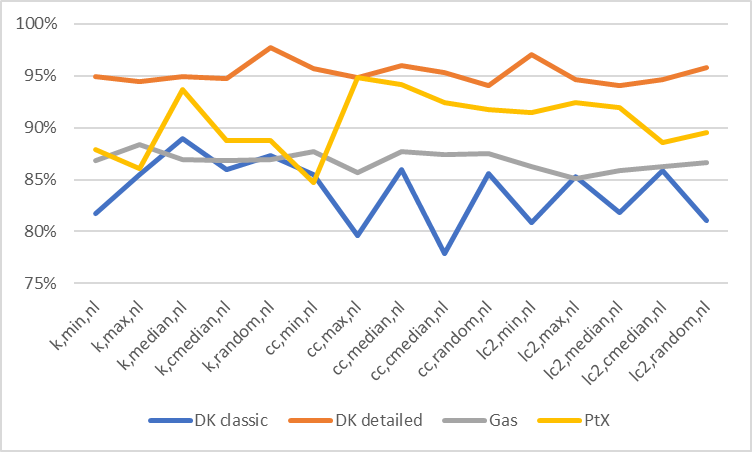}
    \caption{\label{fig11}Time usage reductions in percent for the non-weighted algorithms with more clusters.}
\end{figure}

\begin{figure}[htbp]
    \centering
    \includegraphics[width=0.9\columnwidth]{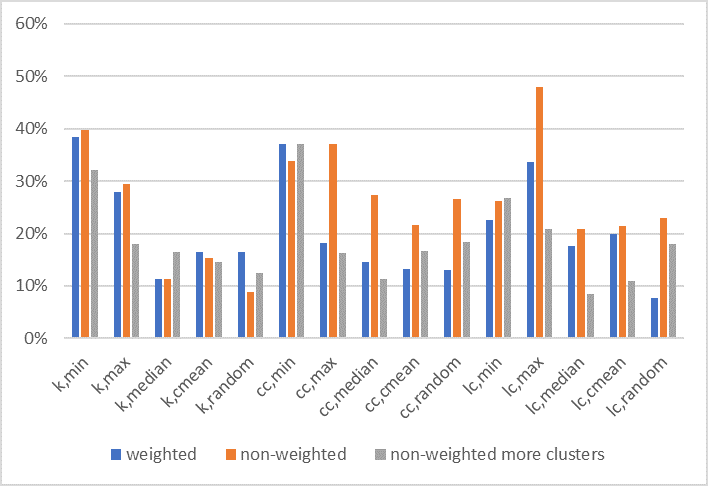}
    \caption{\label{fig12}Quality gap percentages averaged across the four instances. Results are illustrated for weighted selection, non-weighted selection and non-weighted selection with more clusters.}
\end{figure}

Time savings are consistent with previous results, see Table \ref{t10} and
Figure \ref{fig11}. Quality results can be seen in Table \ref{t11} and Figure \ref{fig12}. The
results reveal that weighted selection is best in 41\% of the tests,
non-weighted in 21\% and non-weighted with more clusters in 38\% of
the instances. Overall, weighted selection still has best performance
with respect to quality, though generating more clusters is also a
very reasonable approach.

\begin{table}[htbp]
    \centering
    \small
    \caption{\label{t10} Run times in minutes. nl denotes non-weighted selection with more clusters.}
    \begin{tabular}{l|llll}
	&DK-classic	&DK-detailed	&Gas	&PtX\\\hline\hline
k,min,nl	&6,65	&27,13&	1,54&	1,78\\
k,max,nl	&5,28	&29,68&	1,36&	2,05\\
k,median,nl	&4,02	&27,39&	1,53&	0,93\\
k,cmedian,nl&	5,12&	28,41&	1,54&	1,65\\
k,random,nl	&4,61	&12,27&	1,53&	1,65\\\hline
cc,min,nl	&5,3	&23,22&	1,44&	2,24\\
cc,max,nl	&7,44	&27,72&	1,68&	0,76\\
cc,median,nl&	5,12&	21,79&	1,44&	0,86\\
cc,cmedian,nl&	8,06&	25,25&	1,47&	1,11\\
cc,random,nl&	5,25&	32,05&	1,46&	1,21\\\hline
lc2,min,nl	&6,97	&15,98&	1,61&	1,25\\
lc2,max,nl	&5,36	&28,64&	1,75&	1,12\\
lc2,median,nl&	6,62&	31,73&	1,65&	1,18\\
lc2,cmedian,nl&	5,15&	28,98&	1,61&	1,68\\
lc2,random,nl&	6,89&	22,34&	1,57&	1,54\\\hline
    \end{tabular}
\end{table}

\begin{table}[htbp]
    \centering
    \small
    \caption{\label{t11} Solution quality: the lower percentage, the better performance. nl denotes non-weighted selection with more clusters.}
    \begin{tabular}{l|llll}
	&DK-classic	&DK-detailed	&Gas	&PtX\\\hline\hline
k,min,nl&   	101\%&	5\%&	23\%&	1\%\\
k,max,nl&   	41\%&	10\%&	16\%&	4\%\\
k,median,nl&	41\%&	2\%&	20\%&	3\%\\
k,cmedian,nl&	15\%&	4\%&	37\%&	2\%\\
k,random,nl&	20\%&	3\%&	24\%&	2\%\\\hline
cc,min,nl&  	106\%&	5\%&	35\%&	3\%\\
cc,max,nl&  	22\%&	8\%&	25\%&	10\%\\
cc,median,nl&	9\%&	2\%&	29\%&	5\%\\
cc,cmedian,nl&	19\%&	3\%&	40\%&	4\%\\
cc,random,nl&	33\%&	7\%&	28\%&	5\%\\\hline
lc2,min,nl& 	64\%&	3\%&	40\%&	0\%\\
lc2,max,nl& 	38\%&	15\%&	23\%&	7\%\\
lc2,median,nl&	23\%&	2\%&	7\%&	1\%\\
lc2,cmedian,nl&	19\%&	4\%&	21\%&	1\%\\
lc2,random,nl&	27\%&	3\%&	41\%&	1\%\\\hline
\end{tabular}
\end{table}

\end{document}